\documentclass[fleqn]{article}

\usepackage{amsmath,amssymb}
\usepackage{graphicx}
\usepackage[top=0.75in, bottom=0.75in, left=0.75in, right=0.75in]{geometry}
\usepackage[small]{caption}
\usepackage[noblocks]{authblk}
\usepackage{hyperref}
\usepackage{ulem}
\usepackage{color}

\begin{document}

\title{\sf Heavy-light diquark masses from QCD sum rules and constituent diquark models of tetraquarks}

\author[1]{R.T.\ Kleiv}
\author[1]{T.G.\ Steele}
\author[2]{Ailin\ Zhang}
\author[3]{Ian\ Blokland}

\affil[1]{Department of Physics and Engineering Physics, University of Saskatchewan, Saskatoon, SK, S7N 5E2, Canada}
\affil[2]{Department of Physics, Shanghai University, Shanghai, 200444, China}
\affil[3]{Department of Science, University of Alberta Augustana Campus, Camrose, AB, T4V 2R3, Canada}

\maketitle

\begin{abstract} 


Diquarks with $J^{P}=0^{\pm}\,,1^{\pm}$ containing a heavy  (charm or bottom) quark and a light quark are investigated using QCD Laplace~sum rules. Masses are determined using appropriately constructed gauge invariant correlation functions, including for the first time next-to-leading order perturbative contributions. The $J^P=0^+$ and $1^+$ charm-light diquark masses are respectively found to be $1.86\pm0.05\,{\rm GeV}$ and $1.87\pm0.10\,{\rm GeV}$, while those of the $0^+$ and $1^+$ bottom-light diquarks are both determined to be $5.08\pm0.04\,{\rm GeV}$. The sum rules derived for heavy-light diquarks with negative parity are poorly behaved and do not permit unambiguous mass predictions, in agreement with previous results for negative parity light diquarks. The scalar and axial vector heavy-light diquark masses are degenerate within uncertainty, as expected by heavy quark symmetry considerations. Furthermore, these mass predictions are in good agreement with masses extracted in constituent diquark models 
of the tetraquark candidates $X(3872)$ and $Y_b(10890)$. Thus these results provide QCD support for the interpretation of the $X(3872)$ and $Y_b(10890)$ as $J^{PC}=1^{++}$ tetraquark states composed of diquark clusters. Further implications for tetraquarks among the heavy quarkonium-like XYZ states are discussed. 
\end{abstract}




\section{Introduction}
\label{theIntroduction} 
 

The discovery of the $X(3872)$ by the Belle collaboration \cite{Choi:2003ue} and its subsequent confirmation by the CDF~\cite{Acosta:2003zx}, D\O~\cite{Abazov:2004kp}, BABAR~\cite{Aubert:2004fc} and LHCb~\cite{Aaij:2011sn} collaborations initiated a new era in hadron spectroscopy. Since then, hadrons have been found in the charmonium and bottomonium spectra that are difficult to reconcile as conventional heavy quarkonia. These are called heavy quarkonium-like or XYZ states, and a comprehensive review of the current experimental situation is given in Ref.~\cite{Beringer:1900zz}. The  $X(3872)$ exemplifies the difficulties in interpreting these states: its mass is $M=3871.68\pm0.17\,{\rm MeV}$, its width is $\Gamma<1.2\,{\rm MeV}$~\cite{Beringer:1900zz} and the LHCb collaboration has clearly established that its quantum numbers are $J^{PC}=1^{++}$~\cite{Aaij:2013zoa}. These properties pose problems for a conventional charmonium interpretation of the $X(3872)$~\cite{Swanson:2006st}. Given the proximity of its 
mass to that of $\bar D D^*$, the $X(3872)$ has been widely interpreted as a four-quark molecular state~\cite{Close:2003sg,Voloshin:2003nt,Swanson:2003tb,Tornqvist:2004qy,AlFiky:2005jd,Thomas:2008ja,Liu:2008tn,Lee:2009hy}. 
A complementary interpretation is that the $X(3872)$ is a tetraquark~\cite{Maiani:2004vq,Ebert:2005nc,Matheus:2006xi,Terasaki:2007uv,Dubnicka:2010kz}. In addition to the $X(3872)$, several XYZ states that are four-quark candidates are discussed in Ref.~\cite{Brambilla:2010cs}. 

Molecules and tetraquarks have very different internal quark structures. In the molecular scenario, two color-singlet mesons form a weakly bound conglomerate, whereas in the tetraquark scenario a diquark and anti-diquark form a tightly bound four-quark state. A diquark is a strongly correlated pair of quarks within a hadron (see Ref.~\cite{Anselmino:1992vg} for a review of applications). Because single gluon exchange leads to an attractive interaction between quarks in a color anti-triplet configuration, diquarks are identical to anti-quarks in terms of color. In Ref.~\cite{Jaffe:2004ph} all possible diquark configurations were classified and it was shown that due to spin interactions, the scalar is the most strongly bound, followed by the vector. However, these spin interactions scale as the inverse of the quark mass, and hence scalar and vector diquarks that contain one or more heavy quarks should be degenerate. 

The tetraquark and molecular currents used in QCD sum rule analyses are related through Fierz transformations, leading to ambiguities in their interpretation which can be addressed through the diquark scenario \cite{Zhang:2006xp} (Ref.~\cite{Nielsen:2009uh} provides a review of the numerous QCD sum rule studies of tetraquarks and molecules among the XYZ states). In addition, the renormalization of four-quark operators is complicated by operator mixing~\cite{Narison:1983kn,Jamin:1985su}. Conversely, the renormalization of the diquark operator is multiplicative and has been studied to two-loop order~\cite{Kleiv:2010qk}. For this reason, QCD sum rule studies of diquarks can be extended to higher orders much more easily. The first QCD sum rule studies of diquarks were given in Refs.~\cite{Dosch:1988hu,Jamin:1989hh}, followed by Refs.~\cite{Zhang:2006xp,Wang:2011ab,Wang:2010sh}. The Bethe-Salpeter~\cite{Wang:2005tq,Yu:2006ty}, Dyson-Schwinger~\cite{Maris:2002yu}, and effective field theory approaches~\cite{Kim:2011ut} have also been used to determine diquark masses. Ref.~\cite{Wang:2010sh} used QCD sum rules to investigate heavy-light diquarks with $J^P=0^+$ and $1^+$. In this paper we will build upon previous work by including next-to-leading order perturbative contributions and negative parity diquarks in our analysis. 

Diquarks are clearly not hadrons, thus their masses must be regarded as constituent masses. Constituent diquark models have been used to study tetraquarks among the XYZ states. In Ref.~\cite{Maiani:2004vq} Maiani \textit{et al.} interpret the $X(3872)$ as a tetraquark composed of charm-light diquarks, and using its mass determine both the scalar and vector charm-light constituent diquark masses to be $1.93\,{\rm GeV}$. Ref.~\cite{Faccini:2013lda} points out that the $Z_c^{\pm}(3895)$, which was very recently discovered by the BESIII~\cite{Ablikim:2013mio} collaboration and quickly confirmed by the Belle~\cite{Liu:2013dau} and CLEO~\cite{Xiao:2013iha} collaborations, was predicted in Ref.~\cite{Maiani:2004vq}. The confirmation of this charged charmonium-like state strongly supports the existence of hadrons outside the constituent quark model. Similarly, Ali \textit{et al}.~\cite{Ali:2011ug} interpret the $Y_b(10890)$ discovered by Belle~\cite{Abe:2007tk} as a tetraquark composed of bottom-light diquarks, 
determining the scalar and vector bottom-light diquark masses to be $5.20\,{\rm GeV}$. The charged bottomonium-like states $Z_b^{\pm}(10610)$ and $Z^{\pm}_b(10650)$~\cite{Belle:2011aa} are also suggested to be tetraquarks. Important features of the analyses in Refs.~\cite{Maiani:2004vq,Ali:2011ug} are the use of heavy-light diquarks whose constituent masses are extracted from fits to tetraquark candidates and the equality of scalar and vector heavy-light diquark masses. In this paper we seek to determine if these heavy-light diquark masses are supported by QCD sum rule analyses, thereby providing a QCD-based test of the heavy-light diquark model of tetraquark states. Because our aim is to compare our results with the heavy-light diquark masses determined in Refs.~\cite{Maiani:2004vq,Ali:2011ug}, our focus is on heavy-light diquarks.

The remainder of the paper is organized as follows: in Section~\ref{theCorrelationFunction} we calculate the $J^P=0^{\pm}\,,1^{\pm}$ heavy-light diquark correlation functions, in Section~\ref{theAnalysis} we construct and analyze the corresponding QCD Laplace sum rules, and in Section~\ref{theConclusion} we make concluding remarks and discuss the phenomenological implications of our results.




\section{Heavy-Light Diquark Correlation Function}
\label{theCorrelationFunction} 

The heavy-light diquark correlation function is defined as
\begin{gather}
\Pi\left(Q^2\right) = i \int d^4 x \, e^{i q \cdot x} \langle 0 | T\left[ \right. J_{\alpha}\left(x\right) S_{\alpha\omega} \left[x\,,0\right] J^\dag_{\omega}\left(0\right) \left. \right] |0\rangle \,,
\label{correlation_fcns}
\end{gather}
where $Q^2=-q^2$ is the Euclidean momentum, and $\alpha$, $\omega$ are color indices. The heavy-light diquark currents are
\begin{gather}
J_\alpha = \epsilon_{\alpha\beta\gamma} Q^{T}_{\beta} C\mathcal{O} q_{\gamma} \,,
\label{diquark_currents}
\end{gather}
where $C$ is the charge conjugation operator, $T$ denotes the transpose, $Q$ is a heavy (charm or bottom) quark field, and $q$ is a light quark field~\cite{Dosch:1988hu,Jamin:1989hh}. The Lorentz structures $\mathcal{O}=\gamma_5\,,I\,,\gamma_\mu\,,\gamma_\mu\gamma_5$ respectively couple to scalar $\left(J^P=0^+\right)$, pseudoscalar $\left(0^-\right)$, axial vector $\left(1^+\right)$, and vector $\left(1^-\right)$ heavy-light diquarks. We denote these as $S$, $P$, $A$, and $V$, respectively. The axial vector and vector correlation functions are given by
\begin{gather}
\Pi^{\rm \left(A,V\right)} \left(q\right) = \frac{1}{d-1}\left(\frac{q^\mu q^\nu}{q^2} - g^{\mu\nu} \right) \Pi^{\rm \left(A,V\right)}_{\mu\nu} \left(q\right) \,,
\label{vector_projection}
\end{gather}
where $d$ is the number of spacetime dimensions. Following Refs.~\cite{Dosch:1988hu,Jamin:1989hh,Zhang:2006xp,Wang:2010sh,Wang:2011ab}, the diquark correlation function~\eqref{correlation_fcns} includes a path-ordered exponential, or Schwinger string, defined as
\begin{gather}
\begin{split}
S_{\alpha\omega} \left[x\,,0\right] = P \exp\left[{ig\frac{\lambda^a_{\alpha\omega}}{2}\int_0^x dz^\mu\ A^a_\mu\left(z\right)}\right]
\,,
\label{general_schwinger_string}
\end{split}
\end{gather}
where $P$ denotes path-ordering and $g=\sqrt{4\pi\alpha}$ is the strong coupling. Ref.~\cite{Dosch:1988hu} demonstrated that the correlation function~\eqref{correlation_fcns} is gauge invariant to leading order for light quark currents. We will show that this is also true for heavy-light diquark currents~\eqref{diquark_currents}.

\begin{figure}[htb]
\centering
\includegraphics[scale=0.45]{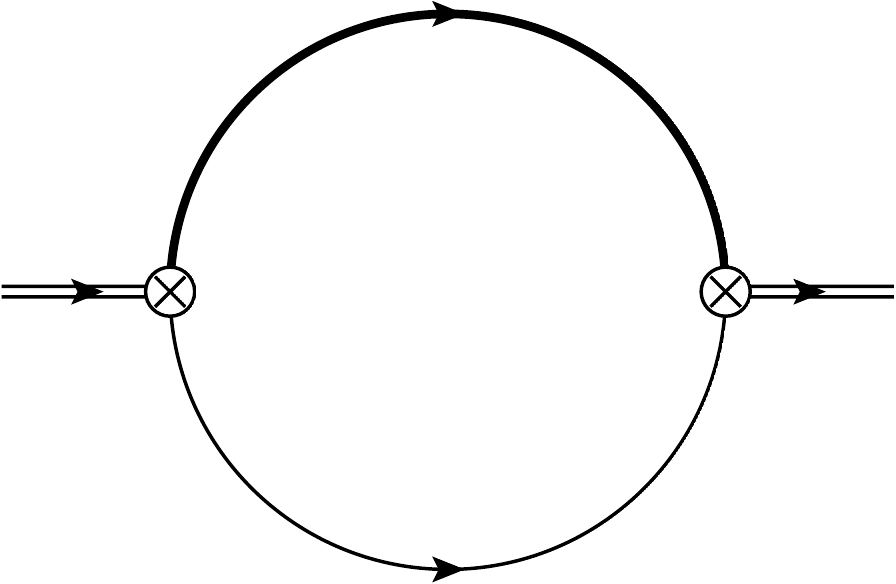}
\includegraphics[scale=0.45]{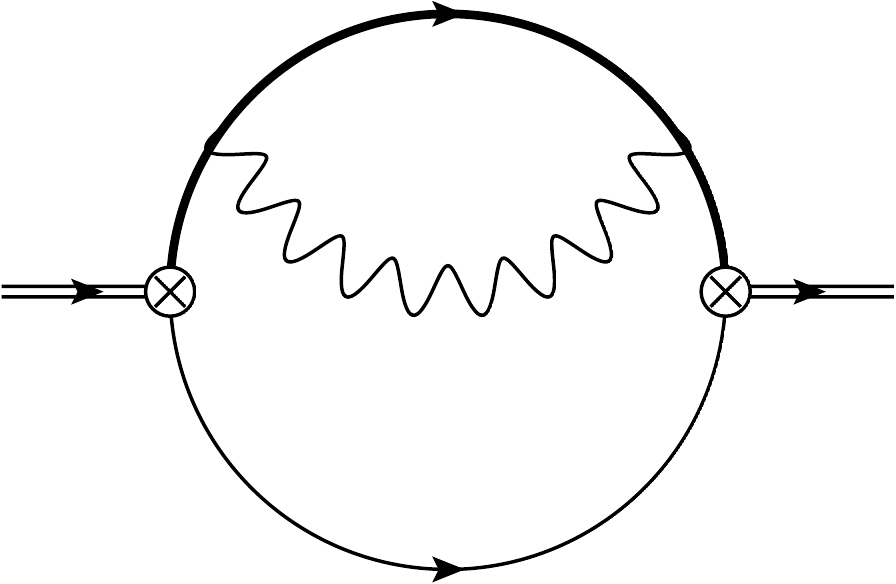}
\includegraphics[scale=0.45]{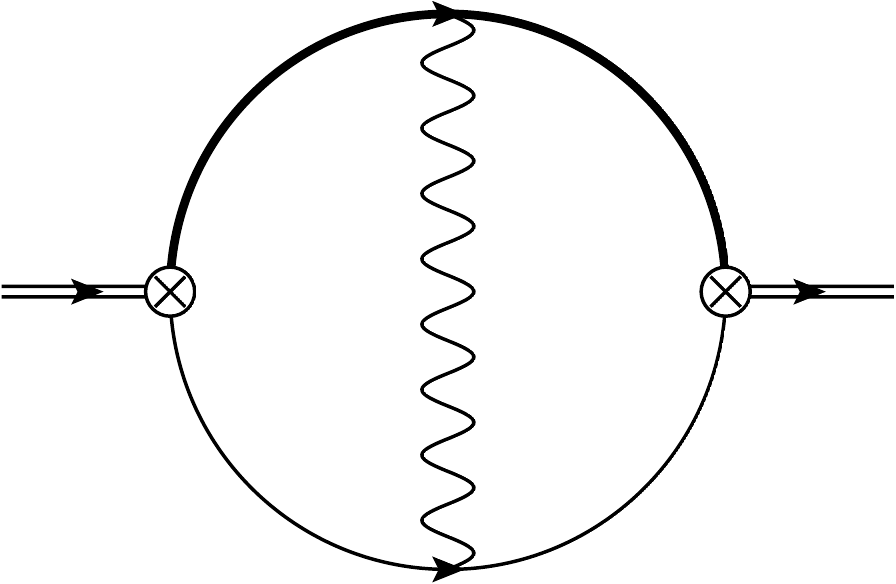}
\caption{
Feynman diagrams representing the leading order and next-to-leading order perturbative contributions to the heavy-light diquark correlation function~\eqref{correlation_fcns}. An insertion of the diquark current is represented by the $\otimes$ symbol, bold lines represent heavy quark propagators, thin lines represent light quark propagators, and wavy lines represent gluon propagators. An additional diagram where the light and heavy quark propagators are exchanged is not shown. These and all subsequent Feynman diagrams were created using JaxoDraw~\cite{Binosi:2003yf}.}
\label{pert_fig}
\end{figure}

First we calculate perturbative contributions to the heavy-light diquark correlation function, which are shown in Fig.~\ref{pert_fig}. We include $\mathcal{O}(\alpha)$ perturbative contributions that have not been calculated previously. To leading order the Schwinger string~\eqref{general_schwinger_string} generates a trace over the color indices in~\eqref{correlation_fcns}, and this trace has been performed in calculating perturbative contributions. We will also consider a higher order contribution from the Schwinger string that is gauge dependent and should cancel the gauge dependence of the perturbative contributions. Thus perturbative contributions are calculated in a general covariant gauge in order to verify the gauge independence of the correlation function~\eqref{correlation_fcns}. The gluon propagator is taken to be
\begin{gather}
D^{AB}_{\mu\nu}(k) = -i\delta^{AB}\left[D^{(0)}_{\mu\nu}(k)-D^{(1)}_{\mu\nu}(k)\right]  \,; \quad D^{(0)}_{\mu\nu}(k)=\frac{g_{\mu\nu}}{k^2} \,, \quad D^{(1)}_{\mu\nu}(k)=(1-a)\frac{k_\mu k_\nu}{k^4} \,,
\label{gluon_prop}
\end{gather}
where $a$ denotes the gauge parameter and the functions $D^{(0)}_{\mu\nu}$, $D^{(1)}_{\mu\nu}$ are defined for later convenience. As in Refs.~\cite{Berg:2012gd,Harnett:2012gs} we calculate the entire correlation function, rather than only the imaginary part. This approach is essential in order to deal with gauge invariance and renormalization issues properly in this calculation. Results for the loop integrals that are encountered are given in Refs.~\cite{Boos:1990rg,Davydychev:1990cq,Pascual_and_Tarrach}. The number of distinct integrals to be calculated can be significantly reduced using the Mathematica package Tarcer~\cite{Mertig:1998vk}, which implements the generalized recurrence relations developed in Refs.~\cite{Tarasov:1996br,Tarasov:1997kx}. Finally, the epsilon expansion can be performed using the Mathematica package HypExp~\cite{Huber:2005yg,Huber:2007dx}. Using the $\overline{\rm MS}$ scheme and working in $d=4+2\epsilon$ dimensions, the 
perturbative result for each channel can be parametrized as 
\begin{gather}
\begin{split}
\Pi^{(i)}_{\rm pert, B}\left(w\right) = \frac{m_B^2}{\pi^2}  & \frac{w+1}{w^2}  \Biggl[  b_0\log{\left(1+w\right)} + \epsilon \left\{b_1\log{\left(1+w\right)}+b_2\log^2{\left(1+w\right)}+b_3{\rm Li}_2\left(\frac{w}{1+w}\right) \right\} \Biggr.
\\
&+\frac{\alpha}{\pi} \left[ \frac{b_4}{\epsilon}\log{\left(1+w\right)} +b_5\log{\left(1+w\right)}{\rm Li}_2\left(\frac{w}{1+w}\right) +b_6\log{\left(1+w\right)} + b_7\log^2{\left(1+w\right)} \right. 
\\
&+ b_8\log^3{\left(1+w\right)} + b_9 {\rm Li}_3\left(-w\right) + b_{10} {\rm Li}_2\left(\frac{w}{1+w}\right) + b_{11} {\rm Li}_3\left(\frac{w}{1+w}\right)
\\
&\Biggl. \left.+a\left\{ \frac{b_{12}}{\epsilon}\log{\left(1+w\right)} +b_{13}\log{\left(1+w\right)} + b_{14}\log^2{\left(1+w\right)} + b_{15} {\rm Li}_2\left(\frac{w}{1+w}\right) \right\} \right] \Biggr] \,, \quad w=\frac{Q^2}{m^2} \,.
\label{bare_pert_result}
\end{split}
\end{gather}
Here the subscript $B$ indicates bare quantities, $i=S\,,\,P\,,\,A\,,\,V$ denotes each distinct channel, ${\rm Li}_3$ and ${\rm Li}_2$ denote the trilogarithm and dilogarithm functions~\cite{LewinPolylogarithms}, and we have omitted terms corresponding to dispersion relation subtraction constants. The coefficients $b_i$ are functions of $w$ which are given for each channel in Table~\ref{bare_pert_coeff_fcns}.

\begin{table}[hbt]
\centering
\begin{tabular}{ lll }
$J^P$ & $0^{\pm}$ & $1^{\pm}$   \\ 
\hline
& & \\
$b_0$ & $\frac{3}{4}w(1+w)$  & $ \frac{1}{4} (1+w) (2 w-1) $  \\
& & \\
$b_1$ & $ \frac{3}{4}w(1+w)\left(L_m-2\right) $ & $ \frac{1}{12} (1+w) \left[8-10 w+(6 w-3) L_m\right] $ \\
& & \\
$b_2$ & $ \frac{3}{8} w (1+w) $ & $ \frac{1}{8} (1+w) (2 w-1) $ \\
& & \\
$b_3$ & $ -\frac{3}{4} w (1+w) $ & $ -\frac{1}{4} (1+w) (2 w-1) $ \\
& & \\
$b_4$ & $ -\frac{3}{4} w (5+w) $ &  $-\frac{3}{2} (w-1) $\\
& & \\
$b_5$ & $ w (1+w) $ & $ \frac{1}{3} (1+w) (2 w-1) $ \\
& & \\
$b_6$ & $ \frac{w}{24} \left[273+87 w+2 \pi ^2 (1+w)-36 (5+w) L_m \right]$&$\frac{1}{36}\left[9w^2+90w-201+\pi ^2 \left(2w^2+w-1\right)-108 (w-1) L_m\right]$ \\
& & \\
$b_7$ & $ -\frac{2+27 w+34 w^2+6 w^3}{8(1+w)} $ & $ \frac{13+2 w-16 w^2}{12(1+w)} $ \\
& & \\
$b_8$ & $ \frac{1}{4} w (1+w) $  & $ \frac{1}{12} (1+w) (2 w-1) $ \\
& & \\
$b_9$ & $ \frac{3}{2} w (1+w) $ & $ \frac{1}{2} (1+w) (2 w-1) $ \\
& & \\
$b_{10}$ & $ \frac{w \left(15+20 w+8 w^2\right)}{4 (1+w)} $ & $ \frac{5w^3+8 w^2-w-9}{6(1+w)} $ \\
& & \\
$b_{11}$ & $ \frac{3}{2} w (1+w) $ & $ \frac{1}{2} (1+w) (2 w-1) $ \\
& & \\
$b_{12}$ & $ \frac{1}{4} w (1+w) $ & $ \frac{1}{12} (1+w) (2 w-1) $ \\
& & \\
$b_{13}$ & $ \frac{1}{8} w \left[4 (1+w) L_m-7-9 w\right] $ & $ \frac{1}{24} \left[11-3 w-16 w^2+4 \left(2w^2+w-1\right) L_m\right]$ \\
& & \\
$b_{14}$ & $ \frac{w \left(3+4 w+2 w^2\right)}{8 (1+w)} $ &  $ \frac{4 w^3+6 w^2-3}{24 (1+w)} $ \\
& & \\
$b_{15}$ & $ -\frac{w \left(1+4 w+2 w^2\right)}{4 (1+w)} $ & $ \frac{1-6 w^2-4 w^3}{12(1+w)} $ \\
& & \\
\hline
\end{tabular}			
\caption{
Coefficient functions $b_i$ for the bare perturbative result~\eqref{bare_pert_result}. Here $L_m = \log{\left[\frac{m^2}{\mu^2}\right]}$.}
\label{bare_pert_coeff_fcns}
\end{table}

Some comments must be made regarding the form of~\eqref{bare_pert_result}. First, terms proportional to the gauge parameter $a$ have been retained to allow comparison with contributions from the path-ordered exponential~\eqref{general_schwinger_string}, so as to ensure that the correlation function~\eqref{correlation_fcns} is gauge invariant. Second, the term $b_4$ in~\eqref{bare_pert_result} is a non-local divergence that cannot be removed through application of the Borel transform when the sum rules are constructed. This term must be dealt with through renormalization, necessitating inclusion of the terms $b_1$, $b_2$, and $b_3$ which will lead to renormalization-induced contributions. A similar methodology was also needed in Ref.~\cite{Harnett:2008cw}. Finally, note that \eqref{bare_pert_result} has a branch cut on $w\in \left(-\infty\,,-1\right]$, as it must. However, after using the package HypExp some functions are generated that do not have this branch structure. This anomalous branch structure is 
spurious and is eliminated when polylogarithm identities are used~\cite{LewinPolylogarithms}.

\begin{figure}[htb]
\centering
\includegraphics[scale=0.45]{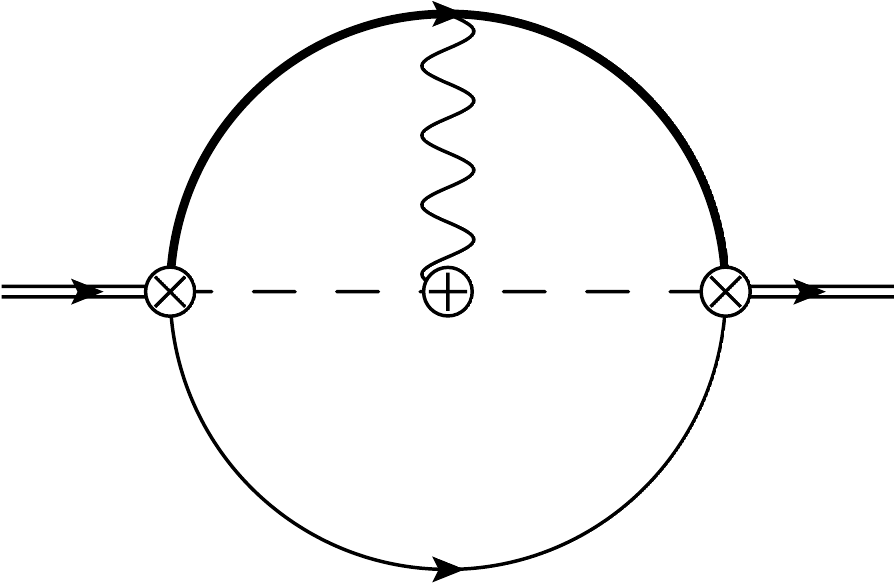}
\caption{
Feynman diagrams representing the contribution of the Schwinger string to the heavy-light diquark correlation function. An insertion of the Schwinger string operator is represented by the $\oplus$ symbol. The dashed line is not a particle propagator. Instead, it indicates the straight line integration path between points $0$ and $x$ used in equation~\eqref{schwinger_string}. An additional diagram where the light and heavy quark propagators are exchanged is not shown. All other notations are identical to Fig.~\ref{pert_fig}.}
\label{string_fig}
\end{figure}

We now turn our attention to contributions from the Schwinger string~\eqref{general_schwinger_string}. Following Ref.~\cite{Dosch:1988hu,Jamin:1989hh}, we define
\begin{gather}
\begin{split}
S_{\alpha\omega} \left[x\,,0\right] = \delta_{\alpha\omega} + ig\frac{\lambda^a_{\alpha\omega}}{2}\int_0^1 d\xi A^a_\mu\left(\xi x\right) x^\mu - g^2 \frac{\lambda^a_{\alpha\beta}}{2} \frac{\lambda^b_{\beta\omega}}{2} \int_0^1 d\xi \int_0^\xi d\xi^\prime :A^a_\mu\left(\xi x\right) A^b_\nu\left(\xi^\prime x\right): x^\mu x^\nu + \mathcal{O}\left(g^3\right) 
\,,
\label{schwinger_string} 
\end{split}
\end{gather}
where $:\phantom{}:$ denotes normal ordering. As in \cite{Dosch:1988hu,Jamin:1989hh} the integration path between points $0$ and $x$ in \eqref{general_schwinger_string} has been chosen to be a straight line.\footnote{In Ref.~\cite{Jamin:1989hh} it was argued that any deviations from a straight line would correspond to additional Wilson loops, and hence would not correspond to the lowest energy configuration.} 
As mentioned earlier, the leading order term in~\eqref{schwinger_string} leads to a trace over the diquark current color indices in~\eqref{correlation_fcns}, which was done in calculating~\eqref{bare_pert_result}. To the order that we are working, the quadratic term in $g$ is irrelevant because it cannot be used to form a gluon propagator. However, the linear term leads to a non-trivial contribution to the correlation function, which is shown in Fig.~\ref{string_fig}. This contribution has the form 
\begin{gather}
\Pi^{(i)}_{\rm string}(Q^2)\sim \int d^4 x \, e^{i q\cdot x} \int d^4 z \, \int_0^1 d\xi \, x^{\mu} \, D^{AB}_{\mu\nu}(\xi x -z)\, \ldots \,,
\label{string_contribution}
\end{gather}
where $q$ is the external momentum, $z$ denotes the location of the quark-gluon interaction in Fig.~\ref{string_fig}, and the ellipses indicate $\xi$-independent terms that are not shown. The $\xi$ integration in \eqref{string_contribution} cannot be evaluated readily. In momentum space the gluon propagator in \eqref{string_contribution} unavoidably leads to terms of the form
\begin{gather}
\Pi^{(i)}_{\rm string}(Q^2) \sim \int \frac{d^d k_1}{(2\pi)^d}\int \frac{d^d k_2}{(2\pi)^d} \int_0^1 d\xi \, D^{AB}_{\mu\nu}(k_1) \frac{\partial }{\partial q_\mu}S\left(q-k_2-\xi k_1 \right)\, \ldots \,,
\label{string_quark_prop}
\end{gather}
where the $\xi$ and loop integrations are coupled. Most of these integrals can be decoupled using the scaling properties of $d$-dimensional momentum integrals~\cite{Collins_Renormalization}, but unfortunately a few cannot be. However, for the gauge dependent terms in~\eqref{string_quark_prop} this obstacle can be circumvented. Note that the quark propagator in~\eqref{string_quark_prop} satisfies the identity 
\begin{gather}
k_1 \cdot \frac{\partial}{\partial q} S\left(q-k_2-\xi k_1 \right) = - \frac{d}{d\xi} S\left(q-k_2-\xi k_1 \right) \,.
\label{xi_identity}
\end{gather}
The $D^{(1)}_{\mu\nu}(k_1)$ part of the gluon propagator~\eqref{gluon_prop} provides a factor of $k_1^\mu$, and hence the $\xi$ integration in~\eqref{string_quark_prop} can be performed using~\eqref{xi_identity}. Note that this approach cannot be used to calculate terms in~\eqref{string_quark_prop} that correspond to the $D^{(0)}_{\mu\nu}(k_1)$ piece of the gluon propagator~\eqref{gluon_prop}. Based upon the result of Ref.~\cite{Dosch:1988hu} we have assumed that there is no contribution from the Schwinger string in Landau gauge. The remaining loop integrations can be performed using the same methods that were used to calculate the perturbative contributions. It should be noted that because of the bosonic nature of the diquark currents and gauge field, the integration over $\xi$ in \eqref{schwinger_string} must be symmetric about the point $\xi=\frac{1}{2}$, meaning that the gauge configurations corresponding to $\xi$ and $1-\xi$ are equivalent. Thus the $\xi$ integration double counts and we have introduced 
an overall factor of $\frac{1}{2}$ accordingly. 

In order to check the validity of the methods described, we have used them to reproduce the result of Ref.~\cite{Dosch:1988hu}, verifying that the correlation function~\eqref{correlation_fcns} for light diquark currents is gauge independent to order $\alpha$. Using the approach described above we have calculated the gauge dependent contributions of the Schwinger string~\eqref{schwinger_string} to the heavy-light diquark correlation function~\eqref{correlation_fcns}. We find that these precisely cancel the gauge dependent terms $b_{12}$, $b_{13}$, $b_{14}$, and $b_{15}$ in the perturbative contribution~\eqref{bare_pert_result}. This verification of gauge independence emerges from the manifestly gauge invariant formalism of the Schwinger string, confirming that the heavy-light diquark correlation function~\eqref{correlation_fcns} is suitable for use in a QCD sum rule analysis.

Now we must renormalize the bare result~\eqref{bare_pert_result}. To the order that we are working, this can be done through renormalization of the heavy quark mass and the diquark current. The one-loop expression for the renormalized quark mass is~\cite{Pascual_and_Tarrach}
\begin{gather}
m_B = Z_m \, m \,, \quad Z_m = 1 + \frac{\alpha}{\pi\epsilon}  \,.
\label{mass_renorm}
\end{gather}
The renormalization of the scalar diquark current was studied in Ref.~\cite{Kleiv:2010qk}. A distinct benefit of using a diquark current rather than a four quark current is that, unlike four quark currents, the diquark current renormalizes multiplicatively. In Ref.~\cite{Kleiv:2010qk} it was shown that the renormalization factors of the scalar diquark and meson operators are proportional at one-loop level. This relationship can be extended to the pseudoscalar, axial vector, and vector channels in order to determine the renormalization factors of those diquark operators. Given our explicit demonstration of gauge independence and that the Schwinger string contributions are zero in Landau gauge~\cite{Dosch:1988hu}, we calculate the renormalization factors in Landau gauge.\footnote{This is the approach that was implicitly used in Refs.~\cite{Dosch:1988hu,Jamin:1989hh}.} The results are as follows:
\begin{gather}
\left[\right. J^{\rm\,(i)}_\alpha\left.\right]_R = Z_d^{\rm\,(i)} \left[\right.J^{\rm\,(i)}_\alpha\left.\right]_B\,; \quad Z_d^{\rm \,(S)} = 1 + \frac{\alpha}{2\pi\epsilon} \,, \quad Z_d^{\rm \,(P)} =  1 + \frac{\alpha}{2\pi\epsilon} \,, \quad Z_d^{\rm \,(A)}=1 \,, \quad Z_d^{\rm\,(V)}=1 \,.
\label{diquark_renorm_factor}
\end{gather}
Note that the axial vector and vector diquark operator renormalization factors are trivial, in analogy with the corresponding meson operators. Finally, the renormalized perturbative result for each distinct heavy-diquark channel can be expressed as
\begin{gather}
\begin{split}
\Pi^{(i)}_{\rm pert}\left(w\right) = \frac{m^2}{\pi^2}   \frac{w+1}{w^2}  \Biggl[ & c_0\log{\left(1+w\right)} + \frac{\alpha}{\pi}  \biggl[ c_1\log{\left(1+w\right)} + c_2\log^2{\left(1+w\right)} + c_3\log^3{\left(1+w\right)} \biggl. \Biggr.
\\
&\Biggl.\biggl. +c_4 \log{\left(1+w\right)} {\rm Li}_{2}\left(\frac{w}{1+w}\right) + c_5 {\rm Li}_{2}\left(\frac{w}{1+w}\right) + c_6 {\rm Li}_{3}\left(-w\right) + c_7 {\rm Li}_{3}\left(\frac{w}{1+w}\right) \biggr] \Biggr] \,.
\label{renorm_pert_result}
\end{split}
\end{gather}
The heavy quark mass and strong coupling are implicitly functions of the renormalization scale $\mu$, and the coefficients $c_i$ are functions of $w$ that are given in Table~\ref{renorm_pert_coeff_fcns}.

\begin{table}[hbt]
\centering
\begin{tabular}{ lll }
$J^P$ & $0^{\pm}$ & $1^{\pm}$   \\ 
\hline
& & \\
$c_0$ & $ \frac{3}{4} w (1+w) $ 		& $ \frac{1}{4} (1+w) (2 w-1) $ 		\\
& & \\
$c_1$ & $\frac{1}{24} w\left[165+51w+2\pi^2(1+w)-18(5+w)L_m\right]$& $ \frac{1}{36}\left[9w^2+90w-93+\pi^2\left(2w^2+w-1\right)-54 (w-1) L_m\right] $ \\
& & \\
$c_2$ & $ -\frac{2+12 w+16 w^2+3 w^3}{8(1+w)} $		& $ \frac{4+2 w-7 w^2}{12(1+w)} $		\\
& & \\
$c_3$ & $ \frac{1}{4} w (1+w) $		& $ \frac{1}{12} (1+w) (2 w-1) $		\\
& & \\
$c_4$ & $ w (1+w) $		& $ \frac{1}{3} (1+w) (2 w-1) $		\\
& & \\
$c_5$ & $ \frac{w^2 (2+5 w)}{4 (1+w)} $		& $ \frac{5 w^3-w^2-w}{6(1+w)} $		\\
& & \\
$c_6$ & $ \frac{3}{2} w (1+w) $ 		& $ \frac{1}{2} (1+w) (2 w-1) $		\\
& & \\
$c_7$ & $ \frac{3}{2} w (1+w) $ 		& $ \frac{1}{2} (1+w) (2 w-1) $		\\
& & \\
\hline
\end{tabular}			
\caption{
Coefficient functions $c_i$ for the renormalized perturbative result~\eqref{renorm_pert_result}. All notations are identical to those in Table~\ref{bare_pert_coeff_fcns}.}
\label{renorm_pert_coeff_fcns}
\end{table}

The imaginary part of~\eqref{renorm_pert_result} can be easily determined via analytic continuation. The result is
\begin{gather}
\begin{split}
{\rm Im}\Pi^{(i)}_{\rm pert}\left(x\right) = \frac{m^2}{4\pi x}  \Biggl[ f_0 + \frac{\alpha}{\pi} \Biggl( f_1 + f_2\log{\left(x\right)} + f_3\log{\left(1-x\right)} + & f_4\log{\left(x\right)}\log{\left(1-x\right)}\Biggr.\Biggr.
\\
&\Biggl.\Biggl.+ f_5{\rm Li}_{2}\left(x\right) +  f_6\log{\left[\frac{m^2}{\mu^2}\right]} \Biggr) \Biggr] \,, \quad 0<x<1 \,,
\label{imaginary_part_of_pert_result}
\end{split}
\end{gather}
where the coefficients $f_i$ are functions of $x$ as given in Table~\ref{Im_pert_coeff_fcns}.

\begin{table}[hbt]
\centering
\begin{tabular}{ lll }
$J^P$ & $0^{\pm}$ & $1^{\pm}$   										\\ 
\hline
& & \\
$f_0$ & $3\left(1-x\right)^2$ 				& $2-3x+x^3$ 						\\
& & \\
$f_1$ & $\frac{1}{2}\left(17-72x+55x^2\right)$		& $\frac{1}{3}\left(3-33x-x^2+31x^3\right)$		\\
& & \\
$f_2$ & $3-16 x+12 x^2-2 x^3$				& $\frac{2}{3} x \left(-7-2 x+4 x^2\right)$		\\
& & \\
$f_3$ & $2\left(x-4\right)\left(1-x\right)^2$		& $-\frac{2}{3}\left(1-x\right)^2\left(5+4x\right)$	\\
& & \\
$f_4$ & $2\left(1-x\right)^2$				& $\frac{2}{3} \left(2-3 x+x^3\right)$			\\
& & \\
$f_5$ & $4\left(1-x\right)^2$				& $\frac{4}{3}\left(2-3x+x^3\right)$			\\
& & \\
$f_6$ & $-3 \left(1-6 x+5 x^2\right)$ 			& $6 x \left(1-x^2\right)$		 		\\
& & \\
\hline
\end{tabular}			
\caption{
Coefficient functions $f_i$ for the imaginary part of the renormalized perturbative result~\eqref{imaginary_part_of_pert_result}.}
\label{Im_pert_coeff_fcns}
\end{table}

\begin{figure}[htb]
\centering
\includegraphics[scale=0.45]{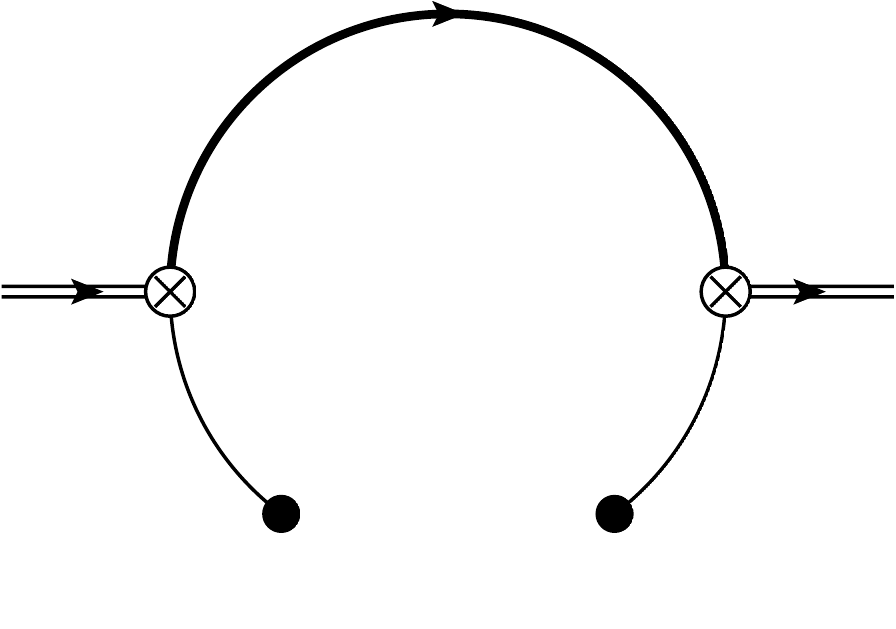}
\caption{
Feynman diagrams representing the dimension-four quark condensate $m_q\langle \bar{q}q \rangle$ contribution to the heavy-light diquark correlation function. Solid dots represent field condensates. All other notations are identical to Fig.~\ref{pert_fig}.}
\label{quark_cond_fig}
\end{figure}

Now we consider contributions to the heavy-light diquark correlation function from the QCD condensates. Following Ref.~\cite{Dosch:1988hu,Jamin:1989hh}, we calculate these contributions using fixed-point gauge techniques because the Schwinger string~\eqref{schwinger_string} does not contribute to the condensates due to the $x^\mu A^a_\mu = 0$ gauge condition. We note that the manifestly gauge invariant nature of the correlation function~\eqref{correlation_fcns} containing the Schwinger string implies that the fixed-point gauge results will be equivalent to those obtained in other methods \cite{Bagan:1992tg}. First we consider the contribution from the quark condensate $\langle \bar{q}q \rangle$, which is shown in Fig.~\ref{quark_cond_fig}. For this contribution we find
\begin{gather}
\Pi^{\rm(S\,,\,A)}_{\rm \bar{q}q}\left(Q^2\right) = -2 \frac{m\langle \bar{q}q \rangle}{Q^2+m^2} \,, \quad \Pi^{\rm (P\,,\,V)}_{\rm \bar{q}q}\left(Q^2\right) =-\Pi^{\rm (S\,,\,A)}_{\rm \bar{q}q}\left(Q^2\right) \,.
\label{quark_cond_result}
\end{gather}
When the QCD Laplace sum rules are constructed in Section~\ref{theAnalysis}, we will need to calculate the Borel transform $\hat B$ of~\eqref{quark_cond_result} and all additional condensate contributions. The following result is useful in order to calculate Borel transforms of the condensate contributions~\cite{Pascual_and_Tarrach}
\begin{gather}
\frac{\hat B}{\tau} \left[\frac{(-Q^2)^k}{Q^2+m^2}\right] = m^{2k} e^{-m^2\tau} \,.
\label{borel_transform}
\end{gather}
This result can be extended to cases where the denominator is raised to a higher power by differentiating~\eqref{borel_transform} with respect to $m^2$. Using this result, the quark condensate contributions to the sum rules are given by
\begin{gather}
\mathcal{B}_{\rm \bar{q}q}^{\rm(S\,,\,A)}\left(k\,,\tau\right) \equiv \frac{\hat B}{\tau} \left[(-Q^2)^k \, \Pi^{\rm(S\,,\,A)}_{\rm \bar{q}q}\left(Q^2\right) \right] = -2 m^{2k} m\langle \bar{q}q \rangle e^{-m^2\tau} \,, 
\quad \mathcal{B}_{\rm \bar{q}q}^{\rm(P\,,\,V)}\left(k\,,\tau\right) = - \mathcal{B}_{\rm \bar{q}q}^{\rm(S\,,\,A)}\left(k\,,\tau\right) \,.
\label{quark_cond_borel_transform}
\end{gather}

\begin{figure}[htb]
\centering
\includegraphics[scale=0.45]{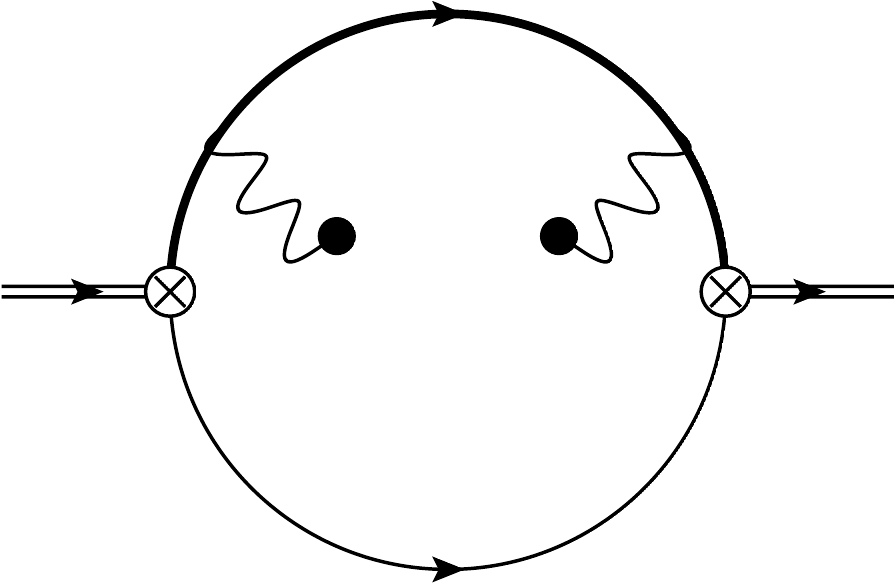}
\includegraphics[scale=0.45]{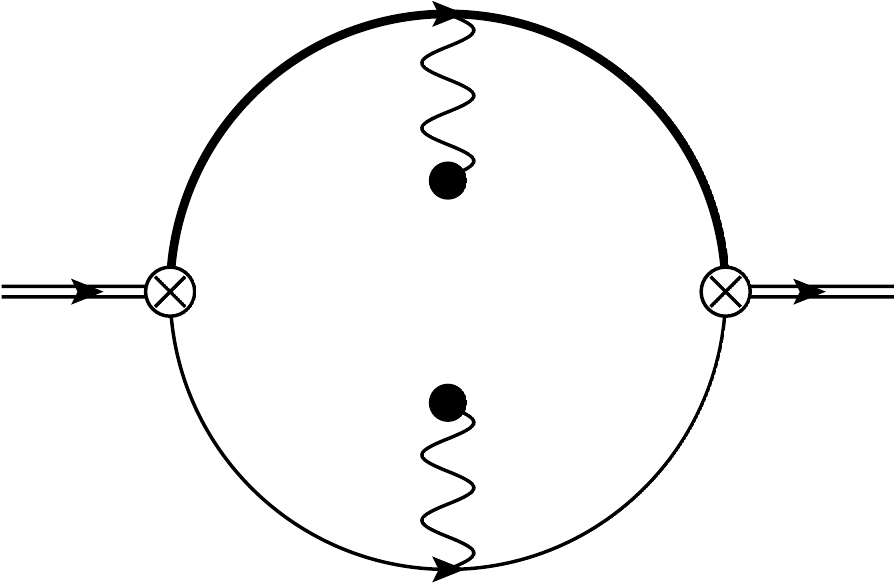}
\caption{
Feynman diagrams representing the dimension-four gluon condensate $\langle \alpha G^2 \rangle$ contribution to the heavy-light diquark correlation function. An additional diagram where the light and heavy quark propagators are exchanged is not shown. All notations are identical to those in Fig.~\ref{quark_cond_fig}.}
\label{gluon_cond_fig}
\end{figure}

Next, we determine contributions from the gluon condensate $\langle \alpha G^2 \rangle=\langle \alpha G^a_{\mu\nu}G_a^{\mu\nu} \rangle$, which are shown in Fig.~\ref{gluon_cond_fig}. For these contributions we find
\begin{gather}
\begin{split}
\Pi^{\rm (S\,,P)}_{\rm GG}\left(Q^2\right) = \frac{\langle \alpha G^2 \rangle}{24\pi} \frac{1}{Q^2+m^2} \,,
\quad
\Pi^{\rm (A\,,V)}_{\rm GG}\left(Q^2\right) = \frac{\langle \alpha G^2 \rangle}{24\pi} \Biggl[ \frac{1}{Q^2}-\frac{3}{Q^2+m^2} 
-\frac{m^2}{Q^4} \log{\left[1+\frac{Q^2}{m^2}\right]} \Biggr] \,.
\end{split}
\label{gluon_cond_result}
\end{gather}
The Borel transforms of these are
\begin{gather}
\begin{split}
\mathcal{B}_{\rm GG}^{\rm(S\,,\,P)}\left(k\,,\tau\right) =  \frac{\langle \alpha G^2 \rangle}{24\pi} m^{2k} e^{-m^2\tau} \,,
\quad 
\mathcal{B}_{\rm GG}^{\rm(A\,,\,V)}\left(k\,,\tau\right) = -\frac{\langle \alpha G^2 \rangle}{8\pi} m^{2k} e^{-m^2\tau} \,.
\end{split}
\label{gluon_cond_borel_transform}
\end{gather}
In calculating~\eqref{gluon_cond_borel_transform} for the axial vector and vector channels we have not included the logarithmic term in~\eqref{gluon_cond_result}. This term will lead to an imaginary part and hence the gluon condensate will have a continuum contribution in these channels. This can be calculated an identical fashion to~\eqref{imaginary_part_of_pert_result}, with the result
\begin{gather}
{\rm Im}\Pi^{\rm (A,V)}_{\rm GG}\left(x\right) = \frac{\langle \alpha G^2 \rangle}{24m^2} x^2 \,, \quad 0<x<1 \,.
\label{gluon_cond_imaginary_part}
\end{gather}

\begin{figure}[htb]
\centering
\includegraphics[scale=0.45]{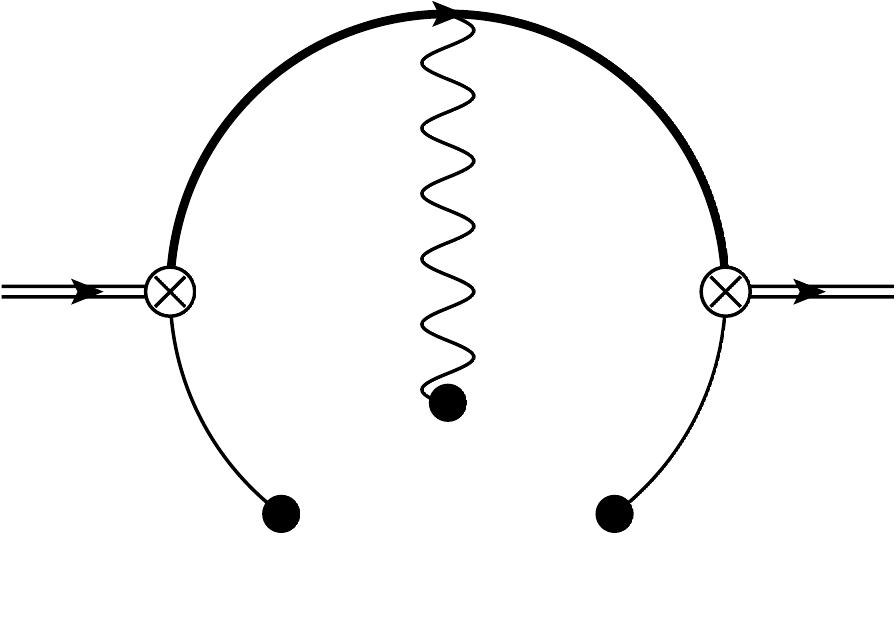}
\caption{
Feynman diagram representing one of the dimension-five mixed condensate $\langle g \bar{q} \sigma G q \rangle $ contributions to heavy-light diquark correlation function. All notations are identical to Fig.~\ref{pert_fig}.}
\label{mixed_cond_fig}
\end{figure}

The contributions of the mixed condensate $\langle \bar{q}\sigma G q \rangle = \langle g \bar{q} \frac{\lambda^a}{2} \sigma^{\mu\nu} G^a_{\mu\nu} q \rangle$ are
\begin{gather}
\begin{split}
&\Pi^{\rm (S)}_{\rm \bar{q}Gq}\left(Q^2\right) = \frac{1}{2} m  \langle \bar{q}\sigma G q \rangle \Biggl[ \frac{m^2 - Q^2 }{\left(Q^2+m^2\right)^3} \Biggr] \,, 
\quad 
\Pi^{\rm (P)}_{\rm \bar{q}Gq}\left(Q^2\right) = -\frac{1}{2} m \langle \bar{q}\sigma G q \rangle \Biggl[ \frac{3m^2 + Q^2 }{\left(Q^2+m^2\right)^3} \Biggr] \,,
\\
&\Pi^{\rm (A)}_{\rm \bar{q}Gq}\left(Q^2\right) \,\, = \,\, m \langle \bar{q}\sigma G q \rangle \Biggl[ \frac{m^2}{\left(Q^2+m^2\right)^3} \Biggr] \,,
\quad \Pi^{\rm (V)}_{\rm \bar{q}Gq}\left(Q^2\right) = - \Pi^{\rm (A)}_{\rm \bar{q}Gq}\left(Q^2\right) \,.
\label{mixed_cond_result}
\end{split}
\end{gather}
Note that~\eqref{mixed_cond_result} includes a term that arises from the fixed-point gauge expansion of the vacuum expectation value $\langle \bar{q}(x) q(0) \rangle$ in Fig.~\ref{quark_cond_fig}. This is separate and distinct from the term that is represented in Fig.~\ref{mixed_cond_fig}. The contributions of the mixed condensate to the sum rules can be calculated using~\eqref{borel_transform}, yielding
\begin{gather}
\begin{split}
&\mathcal{B}_{\rm \bar{q}Gq}^{\rm(S)}\left(k\,,\tau\right) =  \frac{1}{2} m \langle \bar{q}\sigma G q \rangle m^{2(k-1)} e^{-m^2\tau} \left[k^2-2km^2\tau+m^2\tau\left(m^2\tau-1\right)\right]\,,
\\
&\mathcal{B}_{\rm \bar{q}Gq}^{\rm(P)}\left(k\,,\tau\right) =  -\frac{1}{2} m \langle \bar{q}\sigma G q \rangle m^{2(k-1)} e^{-m^2\tau} \left[k^2-2k\left(1+m^2\tau\right)+m^2\tau\left(1+m^2\tau\right)\right]\,,
\\
&\mathcal{B}_{\rm \bar{q}Gq}^{\rm(A)}\left(k\,,\tau\right) =  \frac{1}{2} m \langle \bar{q}\sigma G q \rangle m^{2(k+1)} e^{-m^2\tau} \left[\tau^2-\frac{2k\tau}{m^2}+\frac{k(k-1)}{m^4}\right]\,, 
\quad
\mathcal{B}_{\rm \bar{q}Gq}^{\rm(V)}\left(k\,,\tau\right) = -\mathcal{B}_{\rm \bar{q}Gq}^{\rm(A)}\left(k\,,\tau\right) \,.
\end{split}
\label{mixed_cond_borel_transform}
\end{gather}

Finally we consider the dimension-six quark condensate, $\alpha\langle\bar{q}q\rangle^2$, which arises purely from a higher order term in the fixed-point gauge expansion of the vacuum expectation value $\langle \bar{q}(x) q(0) \rangle$  in Fig.~\ref{quark_cond_fig}. For this we find
\begin{gather}
\begin{split}
&\Pi^{\rm (S)}_{\rm \bar{q}q\bar{q}q }\left(Q^2\right) = -\frac{16\pi}{27} \alpha\langle\bar{q}q\rangle^2  \Biggl[\frac{m^4}{\left(Q^2+m^2\right)^4}\Biggr] \,,
\quad \Pi^{\rm (P)}_{\rm \bar{q}q\bar{q}q }\left(Q^2\right) = \Pi^{\rm (S)}_{\rm \bar{q}q\bar{q}q }\left(Q^2\right) \,,
\\
&\Pi^{\rm (A)}_{\rm \bar{q}q\bar{q}q }\left(Q^2\right) = -\Pi^{\rm (S)}_{\rm \bar{q}q\bar{q}q }\left(Q^2\right) \,,
\quad 
\Pi^{\rm (V)}_{\rm \bar{q}q\bar{q}q }\left(Q^2\right) = -\Pi^{\rm (S)}_{\rm \bar{q}q\bar{q}q }\left(Q^2\right) \,,
\label{dim_six_quark_cond_result}
\end{split}
\end{gather}
where we have assumed vacuum saturation. The contributions of the dimension-six quark condensate to the sum rules are given by
\begin{gather}
\begin{split}
&\mathcal{B}_{\rm \bar{q}q\bar{q}q}^{\rm(S,P)}\left(k\,,\tau\right) = - \frac{8\pi}{81} \alpha\langle \bar{q}q \rangle m^{2(k+2)} e^{-m^2\tau}
\left[\tau^3 - \frac{3k\tau^2}{m^2} +\frac{3k(k-1)\tau}{m^4}-\frac{k(k-1)(k-2)}{m^6}\right] \,,
\\
&\mathcal{B}_{\rm \bar{q}q\bar{q}q}^{\rm(A,V)}\left(k\,,\tau\right) = -\mathcal{B}_{\rm \bar{q}q\bar{q}q}^{\rm(S,P)}\left(k\,,\tau\right) \,.
\end{split}
\label{dim_six_quark_cond_borel_transform}
\end{gather}
We do not consider the dimension-six gluon condensate in this analysis. In Section~\ref{theAnalysis} we will see that the gluon condensate is a sub-leading contribution to the heavy-light diquark sum rules, hence we expect higher-dimensional gluon condensates are suppressed and can be ignored.




\section{QCD Laplace Sum-Rule Analysis}  
\label{theAnalysis}

We now proceed to the QCD Laplace sum rules analysis of $J^P=0^{\pm}\,,1^{\pm}$ heavy-light diquarks. Refs.~\cite{Shifman:1978bx,Shifman:1978by} are the original papers presenting the QCD sum rules technique, and reviews of its methodology are given in Refs.~\cite{Reinders:1984sr,Narison:2002pw}. Using a resonance plus continuum model for the hadronic spectral function
\begin{gather}
\rho^{\rm had}(t) = \rho^{\rm res}(t) + \theta(t-s_0) {\rm Im} \Pi\left(t\right) \,,
\label{hadronic_spectral_fcn}
\end{gather}
where $s_0$ is the continuum threshold, the Laplace sum rules take the form
\begin{equation}
 {\cal R}_{k}\left(\tau,s_0\right)  = \frac{1}{\pi}\int_{t_0}^{\infty} t^k
   \exp\left[ -t\tau\right] \rho^{\rm res}\left(t\right)\; dt \,,
\label{final_laplace}
\end{equation}
where $t_0$ is the hadronic threshold. The left hand side of \eqref{final_laplace} is given by
\begin{equation}
{\cal R}_k\left(\tau,s_0\right)\equiv\frac{\hat B}{\tau}\left[\left(-Q^2\right)^k\Pi\left(Q^2\right)\right] -  \frac{1}{\pi}  \int_{s_0}^{\infty} t^k
   \exp \left[-t\tau  \right]  {\rm Im} \Pi\left(t\right)\; dt  \,.
\label{laplace}
\end{equation}

We now construct the heavy-light diquark sum rules. Using the results obtained above for the perturbative~\eqref{imaginary_part_of_pert_result}, quark condensate~\eqref{quark_cond_borel_transform}, gluon condensate~\eqref{gluon_cond_borel_transform} and \eqref{gluon_cond_imaginary_part}, mixed condensate~\eqref{mixed_cond_borel_transform}, and dimension-six quark condensate~\eqref{dim_six_quark_cond_borel_transform} contributions, the QCD Laplace sum rules are given by
\begin{gather}
\begin{split}
{\cal R}_k^{\rm (i)}\left(\tau,s_0\right)=\frac{m^2}{\pi} & \int_1^{s_0/m^2} \left(m^2z\right)^k \Biggl[   {\rm Im}\Pi^{\rm (i)}_{\rm pert}\left(\frac{1}{z}\right)+{\rm Im}\Pi^{\rm(i)}_{\rm GG}\left(\frac{1}{z}\right)\Biggr]  e^{-m^2 \tau z} dz 
\\
&+ \mathcal{B}^{(i)}_{\rm \bar{q}q}\left(k\,,\tau\right) + \mathcal{B}^{(i)}_{\rm GG}\left(k\,,\tau\right) + \mathcal{B}^{(i)}_{\rm \bar{q}Gq}\left(k\,,\tau\right) + \mathcal{B}^{(i)}_{\rm \bar{q}q\bar{q}q}\left(k\,,\tau\right) \,.
\label{sum_rule_k}
\end{split}
\end{gather}
The mass and coupling in \eqref{sum_rule_k} are implicitly functions of the renormalization scale $\mu$ in the $\overline{\rm MS}$-scheme and renormalization group improvement may be implemented by setting $\mu=1/\sqrt{\tau}$~\cite{Narison:1981ts}. In order to extract mass predictions for heavy-light diquarks we utilize a single narrow resonance model
\begin{equation}
 \frac{1}{\pi}\rho^{\rm res}(t)=f^2\delta\left(t-M^2\right)\,.
 \label{narrow_res}
\end{equation}
Eqn.~\eqref{final_laplace} then yields
\begin{equation}
{\cal R}_k\left(\tau,s_0\right)=f^2 M^{2k}\exp{\left(-M^2\tau\right)}\,,
\label{narrow_sr}
\end{equation}
from which the heavy-light diquark mass $M$ can be determined via the ratio
\begin{equation}
M=\sqrt{\frac{{\cal R}_1\left(\tau,s_0\right)}{{\cal R}_0\left(\tau,s_0\right)}}\,.
\label{ratio}
\end{equation}

Prior to extracting mass predictions we must discuss the QCD parameters occurring in the sum rules. We use one-loop $\overline{\rm MS}$ expressions for the running coupling, charm and bottom quark masses:
\begin{gather}
\alpha(\mu)=\frac{\alpha\left(M\right)}{1+A\frac{\alpha\left(M\right)}{\pi}\log{\left(\frac{\mu^2}{M^2}\right)}} \,, \quad
m(\mu)=\overline{m}\left(\frac{\alpha(\mu)}{\alpha\left(\overline{m}\right)}\right)^{1/A} \,, \quad \overline{m} = m\left(\mu=m\right).
\label{running_quantities}
\end{gather}
In the charm-light diquark analysis we take 
\begin{gather}
\begin{split}
M = M_\tau = 1.77\,{\rm GeV} \,, \quad \alpha(M_\tau) = 0.33\pm0.01\,, \quad A=A_c = \frac{25}{12}\,, \quad \overline{m}_c = 1.28\pm0.03\,{\rm GeV}\,,
\label{charm_running_quantities}
\end{split}
\end{gather}
while in the bottom-light diquark analysis we use
\begin{gather}
\begin{split}
M = M_Z = 91.188\,{\rm GeV} \,, \quad \alpha(M_Z) =0.1184\pm0.0007 \,, \quad A=A_b = \frac{23}{12}\,, \quad \overline{m}_b= 4.18\pm0.03\,{\rm GeV}\,.
\label{bottom_running_quantities}
\end{split}
\end{gather}
All of these parameters are taken from Ref.~\cite{Beringer:1900zz}, apart from $A_c$ and $A_b$ which are given in Ref.~\cite{Pascual_and_Tarrach}. We set $\mu=1/\sqrt{\tau}$ in order to implement renormalization group improvement as described above. 

We now specify the values used for the QCD condensates. Beginning with the quark condensate, we define
\begin{gather}
m\langle \bar{q}q \rangle = \frac{m\left(2\,{\rm GeV}\right)}{m_q\left(2\,{\rm GeV}\right)} m_q \langle \bar{q}q \rangle \,,
\label{quark_condensate} 
\end{gather}
where $m$ denotes the charm or bottom quark mass and we use the PCAC relation $m_q \langle \bar{q}q \rangle=-\frac{1}{2} f_\pi^2 m_\pi^2$. The numerical values are again taken from Ref.~\cite{Beringer:1900zz}:
\begin{gather}
m_q\left(2\,{\rm GeV}\right) = \frac{1}{2}\left[m_u\left(2\,{\rm GeV}\right)+m_d\left(2\,{\rm GeV}\right)\right] = 0.0038\pm0.0006\,{\rm GeV}\,, \quad f_\pi = 0.093\,{\rm GeV}\,, \quad m_\pi = 0.139\,{\rm GeV} \,, 
\label{pion_and_light_quark_data} 
\\
r_c = \frac{m_c\left(2\,{\rm GeV}\right)}{m_q\left(2\,{\rm GeV}\right)} = 305 \pm 59 \,,
\quad 
r_b = \frac{m_b\left(2\,{\rm GeV}\right)}{m_q\left(2\,{\rm GeV}\right)} = 1229 \pm 210 \,,
\label{2_GeV_ratios}
\end{gather}
where the heavy quark mass at 2 GeV is determined using \eqref{running_quantities}. The mixed condensate is similarly defined as
\begin{gather}
m \langle \bar{q}\sigma G q \rangle = M_0^2 \, m\langle \bar{q}q \rangle \,,
\label{mixed_condensate}
\end{gather}
where $M_0^2=\left(0.8\pm0.1\right)\,{\rm GeV}^2$~\cite{Dosch:1988vv} and $m \langle \bar{q}q \rangle$ is as defined in \eqref{quark_condensate}. The gluon condensate is taken to be
\begin{gather}
\langle \alpha G^2\rangle=\left(7.5\pm 2.0\right)\times 10^{-2}\,{\rm GeV^4}~ \cite{Narison:2010cg}\,.
\label{GG_value}
\end{gather}
Finally, the dimension-six quark condensate is 
\begin{gather}
\alpha\langle \bar{q}q \rangle^2 = (5.8\pm0.9) \times 10^{-4}\,{\rm GeV}^6 \,,
\label{dim_six_quark_cond}
\end{gather}
which implicitly includes deviation from ideal vacuum saturation~\cite{Narison:2005zg}. In condensate contributions there are additional factors of the quark mass that are not included in the definitions~\eqref{quark_condensate} or~\eqref{mixed_condensate}, such as the factors of $m^{2k}$ in~\eqref{quark_cond_borel_transform}, for instance. We define these masses in terms of the pole mass following the approach of Ref.~\cite{Narison:2012xy}, utilizing the known relationship between the pole mass and $\overline{\rm MS}$ mass~\cite{Gray:1990yh,Broadhurst:1991fy,Fleischer:1998dw,Chetyrkin:1999qi}:
\begin{gather}
m = m\left(\mu\right)\left[1+\left(\frac{4}{3}-\log{\left[\frac{\overline{m}^2}{\mu^2}\right]}\right)\frac{\alpha\left(\mu\right)}{\pi}\right] \,,
\label{pole_mass}
\end{gather}
where $m\left(\mu\right)$ and $\alpha\left(\mu\right)$ are determined via~\eqref{running_quantities} and $\overline{m}$ is the one-loop $\overline{\rm MS}$ charm or bottom quark mass.

In order to extract mass prediction for heavy-light diquarks using~\eqref{ratio} we must first establish a permissible range of values for the Borel scale $\tau$ and the continuum threshold $s_0$. We adopt the approach developed in Ref.~\cite{Benmerrouche:1995qa}, whereby the H\"older inequalities~\cite{Beckenbach:1961,Berberian:1965} 
\begin{gather}
\begin{split}
&\left| \int_{t_1}^{t_2} f\left(t\right) g\left(t\right) d\mu \right| \leq \left[ \int_{t_1}^{t_2} \left|f\left(t\right)\right|^p d\mu \right]^{1/p} \left[ \int_{t_1}^{t_2} \left|g\left(t\right)\right|^q d\mu \right]^{1/q} \,, \quad \frac{1}{p}+\frac{1}{q} = 1 \,, \quad p\,,q \geq 1\,,
\label{holder_definition}
\end{split}
\end{gather}
are used to constrain the values of $\tau$ and $s_0$. The key observation of Ref.~\cite{Benmerrouche:1995qa} is that because ${\rm Im}\Pi\left(Q^2\right)$ is related to a physical hadronic spectral function via duality, ${\rm Im}\Pi\left(Q^2\right)$ must be positive and hence it can serve as the integration measure in~\eqref{holder_definition}. It can be shown that the sum rules~\eqref{sum_rule_k} must satisfy
\begin{gather}
\frac{\mathcal{R}_2\left(\tau\,,s_0\right)/\mathcal{R}_1\left(\tau\,,s_0\right)}{\mathcal{R}_1\left(\tau\,,s_0\right)/\mathcal{R}_0\left(\tau\,,s_0\right)} \geq 1 \,, \quad \frac{\mathcal{R}_3\left(\tau\,,s_0\right)/\mathcal{R}_2\left(\tau\,,s_0\right)}{\mathcal{R}_2\left(\tau\,,s_0\right)/\mathcal{R}_1\left(\tau\,,s_0\right)} \geq 1 \,,
\label{holder_ratios}
\end{gather}
where the first and second inequalities come from requiring that $\mathcal{R}_0\left(\tau\,,s_0\right)$ and $\mathcal{R}_1\left(\tau\,,s_0\right)$ satisfy the H\"older inequalities, respectively. The inequalities in~\eqref{holder_ratios} can be used to set a lower bound on the Borel mass $M_{B} = 1/\sqrt{\tau}$ or to set a lower bound on the continuum threshold $s_0$. The constraints set by the first inequality in~\eqref{holder_ratios} are more restrictive than those set by the second, hence we rely solely upon the first. We fix an upper bound on $M_B$ by requiring that continuum contributions are less than 50\% of total contributions to the sum rule~\cite{Shifman:1978by}
\begin{gather}
f_{\rm cont}\left( \tau,s_0 \right)=\frac{ {\cal R}_1\left(\tau,s_0\right)/{\cal R}_0\left(\tau,s_0\right) }{{\cal R}_1\left(\tau,\infty\right)/{\cal R}_0\left(\tau,\infty\right) }
\label{f_cont}
\end{gather}
and require that $f_{\rm cont} \geq 0.5$. Using~\eqref{holder_ratios} and~\eqref{f_cont} we can define a range of $M_B$ values over which the sum rule is considered reliable, \textit{i.e.}, the sum rule window. We also require that the mass prediction $M\left(\tau\,,s_0\right)$ extracted from~\eqref{ratio} exhibits $\tau$ stability, that is,
\begin{gather}
\frac{d}{d\tau} M \left(\tau\,,s_0\right) = 0
\label{tau_stability} 
\end{gather}
within the sum rule window. However, note that the bounds on the Borel scale that are determined using~\eqref{holder_ratios} and~\eqref{f_cont} will vary depending on the value of $s_0$. Typically the sum rule window widens as $s_0$ is increased. Thus we first seek a minimum value $s_0^{\rm min}$, which we take to be the smallest value of $s_0$ in whose sum rule window $\tau$ stability~\eqref{tau_stability} is satisfied. If there are no values of $s_0$ that exhibit $\tau$ stability, we consider the sum rule to be unstable. Once the minimum value of $s_0$ has been found, we determine the optimal value $s_0^{\rm opt}$ using the following criterion:
\begin{gather}
\chi^2\left(s_0\right)=\sum_j \left( \frac{1}{M}\sqrt{\frac{{\cal R}_1\left(\tau_j,s_0\right)}{{\cal R}_0\left(\tau_j,s_0\right)}}-1 \right)^2\,, \quad  s_0 \geq s_0^{\rm min} \,.
\label{chi_squared}
\end{gather}
The optimal value $s_0^{\rm opt}$ is that for which~\eqref{chi_squared} is minimized. We adopt a conservative approach where~\eqref{chi_squared} is calculated over the sum rule window corresponding to the minimal value of $s_0$. In some cases we can obtain an upper bound on the mass prediction~\eqref{ratio} by taking $s_0 \to \infty$, however in order to extract such a bound the requirements described above must be satisfied.

\begin{table}[htb]
\centering
\begin{tabular}{ cccccccc }
$\left[Qq\right]$ & $J^P$ & $M\,{\rm \left(GeV\right)}$ & $M_{\rm max}\,{\rm \left(GeV\right)}$ & $s_0^{\rm min}\,{\rm \left(GeV^2\right)}$ & $M_B^{\rm min}\,{\rm \left(GeV\right)}$ & $M_B^{\rm max}\,{\rm \left(GeV\right)}$ & $s_0^{\rm opt}\,{\rm \left(GeV^2\right)}$  \\ 
\hline
& & & & & & & \\
$\left[cq\right]$	& $0^+$ & $1.86\pm0.05$ & 2.02 & 5.0 & 1.2 & 1.6 & 5.0 \\
			& $1^+$ & $1.87\pm0.10$ & 2.07 & 5.0 & 1.3 & 1.6 & 5.0 \\
& & & & & & & \\
$\left[bq\right]$	& $0^+$ & $5.08\pm0.04$ & 5.32 & 30 & 2.1 & 3.8 & 30 \\
			& $1^+$ & $5.08\pm0.04$ & 5.32 & 30 & 2.2 & 3.8 & 30 \\
\hline
\end{tabular}			
\caption{
Mass predictions and sum rule parameters for charm-light ($\left[cq\right]$) and bottom-light ($\left[bq\right]$) diquarks with positive parity. $M_{\rm max}$ is an upper bound on the mass, determined from $s_0 \to \infty$. The minimal value of the continuum threshold is $s_0^{\rm min}$,  the optimal value determined by~\eqref{chi_squared} is $s_0^{\rm opt}$, the sum rule window boundaries are $M_B^{\rm min}$ and $M_B^{\rm max}$.}
\label{mass_predictions}
\end{table}

\begin{figure}[hbt]
\centering
\includegraphics[scale=1]{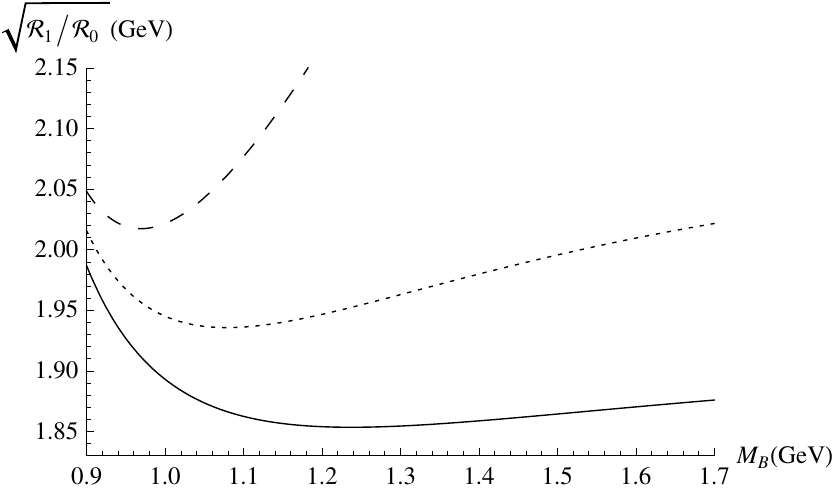}
\includegraphics[scale=1]{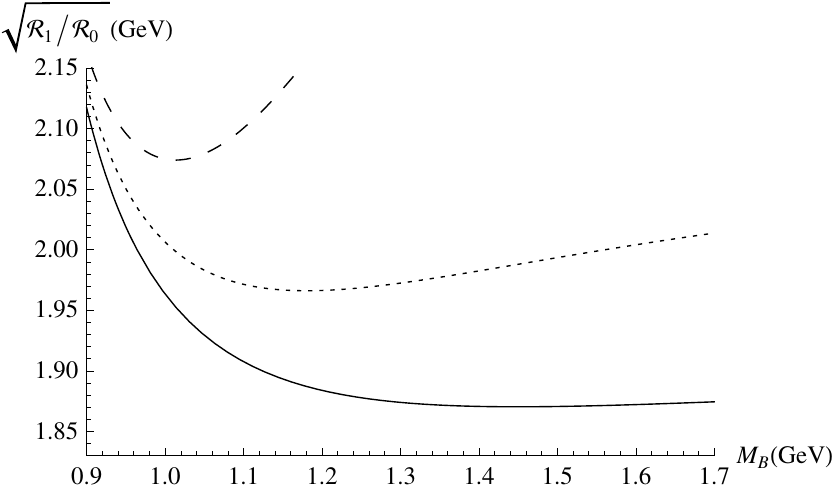}
\caption{
Mass predictions for $J^P=0^+$ (left) and $1^+$ (right) charm-light diquarks. Solid lines correspond to $s_0^{\rm opt}$, yielding the results in Table~\ref{mass_predictions}. In both plots the uppermost dashed lines correspond to $s_0\to\infty$ which provides an upper mass bound and the middle dotted line corresponds to $s_0=6.0\,{\rm GeV}^2$.}
\label{Charm_Scalar_Axial_Results}
\end{figure}

\begin{figure}[hbt]
\centering
\includegraphics[scale=1]{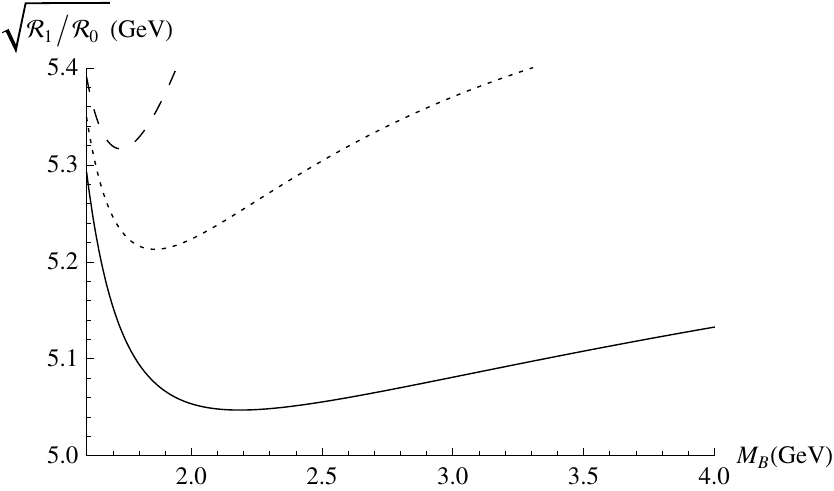}
\includegraphics[scale=1]{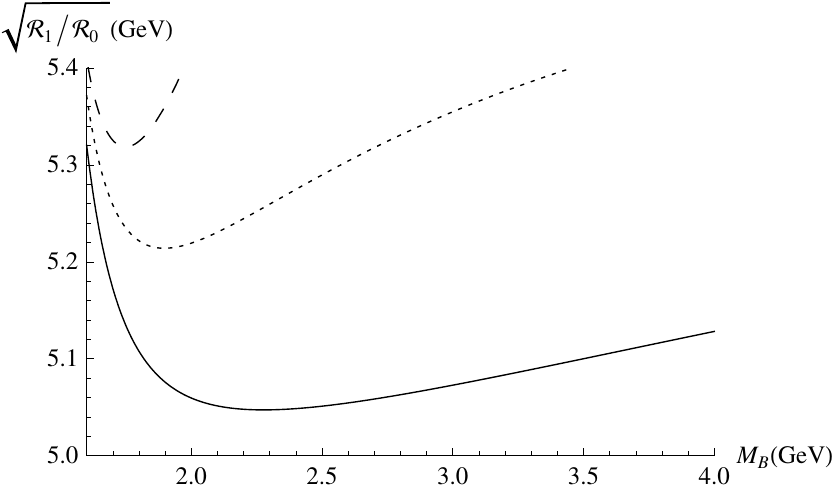}
\caption{
Mass predictions for $J^P=0^+$ (left) and $1^+$ (right) bottom-light diquarks. Solid lines correspond to $s_0^{\rm opt}$, yielding the results in Table~\ref{mass_predictions}. In both plots the uppermost dashed lines correspond to $s_0\to\infty$ and the middle dotted line corresponds to $s_0=35\,{\rm GeV}^2$.
}
\label{Bottom_Scalar_Axial_Results}
\end{figure}

We also determine the uncertainty in our mass predictions due to uncertainties in the QCD parameters. In order of significance these are $r_c$~\eqref{2_GeV_ratios}, $\overline{m}_c$~\eqref{charm_running_quantities}, $\alpha\left(M_\tau\right)$~\eqref{charm_running_quantities}, and $M_0^2$~\eqref{mixed_condensate} in the charm analysis, whereas in the bottom analysis they are $r_b$~\eqref{2_GeV_ratios}, $\overline{m}_b$~\eqref{charm_running_quantities}, $M_0^2$~\eqref{mixed_condensate}, and $\alpha\left(M_Z\right)$~\eqref{charm_running_quantities}. Uncertainties in $\langle \alpha G^2 \rangle$ and $\alpha \langle \bar{q}q \rangle^2$ are insignificant in both cases and we have made no attempt to estimate contributions to these uncertainties from higher loop effects. The resulting mass predictions and uncertainties for heavy-light diquarks with positive parity are summarized in Table~\ref{mass_predictions}. None of the negative parity heavy-light diquark sum rules exhibit $\tau$ stability, therefore we have 
been 
unable to extract mass predictions in these channels. Fig.~\ref{Charm_Scalar_Axial_Results} shows the mass predictions for $J^P=0^+$ and $1^+$ charm-light diquarks while those for bottom-light diquarks are shown in Fig.~\ref{Bottom_Scalar_Axial_Results}.

\section{Conclusions}
\label{theConclusion}

In this paper we have used QCD Laplace sum rules to study heavy-light diquarks with $J^P=0^\pm\,,1^\pm$. Our calculations extend previous sum rule work~\cite{Wang:2010sh} by including higher-loop perturbative contributions which necessitate renormalization of the diquark currents. We have successfully extracted mass predictions for positive parity charm-light and bottom-light diquarks, which are summarized in Table~\ref{mass_predictions}. However, the sum rules derived for negative parity channels are poorly behaved, and do not permit unambiguous mass predictions, similar to what was found for light diquarks~\cite{Jamin:1989hh}. 

The mass predictions for the $J^P=0^+$ and $1^+$ heavy-light diquarks are degenerate within uncertainty, as would be expected by heavy-quark symmetry~\cite{Maiani:2004vq}. Our predicted $J^P=0^+$ and $1^+$ charm-light diquark masses of $1.86\pm0.05\,{\rm GeV}$ and $1.87\pm0.10\,{\rm GeV}$ are in superb agreement with the constituent charm-light diquark mass of $1.93\,{\rm GeV}$ determined by Maiani \textit{et al.}~\cite{Maiani:2004vq} from a fit to the $X(3872)$. Additionally, we predict both the $J^P=0^+$ and $1^+$ bottom-light diquark masses to be $5.08\pm0.04\,{\rm GeV}$ in reasonable agreement with the constituent bottom-light diquark mass of $5.20\,{\rm GeV}$ determined by Ali \textit{et al.}~\cite{Ali:2011ug} from a fit to the $Y_b(10890)$. Given the agreement between these constituent diquark masses and our QCD-based calculations, our results provide QCD support for the identification of the $X(3872)$ and $Y_b(10890)$ as $J^{PC}=1^{++}$ tetraquarks  composed of diquark clusters. Furthermore, because 
the constituent heavy-light diquark is such an important input for constituent diquark models of tetraquarks, we interpret this agreement as indirect support for the predictions of these models. Specifically, our results strengthen the case for the tetraquark interpretation of the charged XYZ states $Z^{\pm}_c(3895)$, $Z_b^{\pm}(10610)$ and $Z_b^{\pm}(10650)$.

In this work we have focused on heavy-light diquarks so as to examine the constituent diquark masses determined in Refs.~\cite{Maiani:2004vq,Ali:2011ug}. However, the methods used in this paper could be extended to doubly-heavy diquarks to study diquark clustering within other tetraquarks or within heavy baryons.

\bigskip
\noindent
{\bf Acknowledgements:}  
TGS and RTK are grateful for the hospitality of Shanghai University where this work was initiated and partially completed. TGS is grateful for financial support from the Natural Sciences and Engineering Research Council of Canada (NSERC). Ailin Zhang is supported by the National Natural Science Foundation of China (11075102) and the Innovation Program of Shanghai Municipal Education Commission under grant No. 13ZZ066.




%

\end{document}